\documentclass[preprint2,times,tighten]{aastex6}
\usepackage{amsmath,amstext}
\usepackage[all]{hypcap} 

\newcommand{\given}{\,|\,}

\newcommand{\dd}{\mathrm{d}}
\newcommand{\bs}[1]{\boldsymbol{#1}}

\newcommand{\plzpars}{\bs{\theta}_{\rm PLZ}}
\newcommand{\tgaspars}{\bs{\theta}_\varpi}
\newcommand{\starpars}{\bs{\alpha}_k}

\shorttitle{Probabilistic Fitting of PLZ Relations}
\shortauthors{Sesar et al.}

\begin{document}

\title{A Probabilistic Approach to Fitting Period-Luminosity Relations and Validating {\em Gaia} Parallaxes}
\author{Branimir Sesar\altaffilmark{1}, Morgan Fouesneau\altaffilmark{1}, Adrian M.~Price-Whelan\altaffilmark{2}, Coryn A.~L.~Bailer-Jones\altaffilmark{1}, Andy Gould\altaffilmark{1}, and Hans-Walter Rix\altaffilmark{1}}
\email{bsesar@mpia.de}
\altaffiltext{1}{Max Planck Institute for Astronomy, K\"{o}nigstuhl 17, D-69117 Heidelberg, Germany}
\altaffiltext{2}{Department of Astrophysical Sciences, Princeton University, Princeton, NJ 08544, USA}

\begin{abstract}
Pulsating stars, such as Cepheids, Miras, and RR Lyrae stars, are important distance indicators and calibrators of the ``cosmic distance ladder'', and yet their period-luminosity-metallicity (PLZ) relations are still constrained using simple statistical methods that cannot take full advantage of available data. To enable optimal usage of data provided by the {\em Gaia} mission, we present a probabilistic approach that simultaneously constrains parameters of PLZ relations and uncertainties in {\em Gaia} parallax measurements. We demonstrate this approach by constraining PLZ relations of type $ab$ RR Lyrae stars in near-infrared W1 and W2 bands, using Tycho-Gaia Astrometric Solution (TGAS) parallax measurements for a sample of $\approx100$ type $ab$ RR Lyrae stars located within 2.5 kpc of the Sun. The fitted PLZ relations are consistent with previous studies, and in combination with other data, deliver distances precise to 6\% (once various sources of uncertainty are taken into account). To a precision of 0.05 mas ($1\sigma$), we do not find a statistically significant offset in TGAS parallaxes for this sample of distant RR Lyrae stars (median parallax of 0.8 mas and distance of 1.4 kpc). With only minor modifications, our probabilistic approach can be used to constrain PLZ relations of other pulsating stars, and we intend to apply it to Cepheid and Mira stars in the near future.
\end{abstract}

\keywords{methods: data analysis --- methods: statistical --- parallaxes  --- stars: variables: RR Lyrae}

\maketitle

\section{Introduction}\label{Sec:Introduction}

One of the main goals of observational astronomy is to ever more precisely and accurately measure the distance to astrophysical objects. Measuring distances to individual stars is critical to understanding a wide range of astronomical phenomena, from stellar structure to Galactic dynamics. An important tool in this endeavor are periodically-pulsating stars, such as Cepheids, Miras, and RR Lyrae stars, whose absolute magnitudes can be predicted using a period-luminosity (PL) relation\footnote{The period-luminosity relation for Cepheids is also known as the Leavitt law \citep{lea08,lea12}.}, and whose period-luminosity relations can be calibrated using trigonometric parallax measurements \citep[e.g., ][]{fc97,vle97,ben07,ben11}. These stars and their PL relations are crucial, as they tie the extragalactic distance scale to the local one, and make the important second rung in the ``cosmic distance ladder''.

The trigonometric parallaxes ($\varpi$) obtained by the Tycho-Gaia Astrometric Solution \citep[TGAS;][]{mic15}, and made public through the Gaia Data Release 1 \citep{gdr1,gdr2}, present us with an exciting opportunity to recalibrate PL relations and potentially improve the accuracy of the cosmic distance ladder. However, the majority of Cepheids, Miras, and RR Lyrae stars in the TGAS sample are distant, and consequently, their parallaxes have a high fractional uncertainty ($\sigma_{\rm \varpi}/\varpi>0.2$). When a parallax has a high fractional uncertainty, the posterior probability distribution over its true distance is complex and non-Gaussian, even when the uncertainties in parallax are Gaussian \citep{b-j15,abj16}. This poses a serious problem for the traditional approach to PL relation fitting, which assumes Gaussian uncertainties in distance, and does a weighted least-squares fit in the distance (i.e., absolute magnitude) vs.~period plane. We can always, of course, fit PL relations by using only stars with precise parallaxes (and thus precise distances with close to Gaussian errors), but then we would ignore a large amount of potentially useful data (that was also obtained at a significant cost).

To avoid ignoring valuable data, in this paper we present a probabilistic approach to inferring period-luminosity(-metallicity) relations that makes a full use of parallax and other measurements, irrespective of their precision. To demonstrate the approach in practice, we use it to constrain period-luminosity-metallicity (PLZ) relations for type $ab$ RR Lyrae stars (i.e., fundamental mode pulsators; \citealt{smi04}) in the near-infrared W1 and W2 bands used by the WISE mission \citep{wri10}. While the demonstration is done using type $ab$ RR Lyrae stars (hereafter, RRab), we emphasize that the approach can be easily applied to other pulsating stars, by simply plugging different data and prior information into our framework.

In addition to constraining PLZ relations, our approach enables a straightforward validation of parallax measurements and their uncertainties. The validation of parallaxes is important as several studies have reported an offset in TGAS parallaxes \citep{st16,jao16} and an overestimation of TGAS parallax uncertainties \citep{cas16,gks16}.

Our paper is organized as follows. In \autoref{Sec:Data}, we describe the astrometric, photometric, and spectroscopic data used to calibrate PLZ relations for RRab stars in WISE W1 and W2 bands, and validate TGAS parallaxes. The construction of the likelihood function that is at the core of our probabilistic approach, is presented in \autoref{Sec:Method}. In \autoref{Sec:Results}, we present the results of applying the probabilistic method of \autoref{Sec:Method}, to data described in \autoref{Sec:Data}. We summarize and discuss our results in \autoref{Sec:Conclusions}.

\section{Data}\label{Sec:Data}

To constrain the PLZ relations for RR Lyrae stars in WISE W1 and W2 bands, we use TGAS trigonometric parallaxes ($\varpi$), spectroscopic metallicities \citep[${\rm [Fe/H]}$; ][]{fer98}, log-periods ($\log P$, base 10), and apparent magnitudes\footnote{Averaged in flux over a pulsation period.} \citep[$m$; ][]{kle14} for 102 RRab stars within $\approx2.5$ kpc from the Sun. The $E(B-V)$ reddening at a star's position is obtained from the \citet{SFD98} dust map. We denote this data set as $\mathcal{D}=\{{\bf d_k}\}$, where ${\bf d_k}=\{\varpi, {\rm [Fe/H]}, \log P, m, EBV\}$ is the data set associated with the $k$th star. 

We note that the extinction correction changes by less than 0.01 mag for 96\% of objects (0.01 to 0.02 mag for the remaining 4\%), if instead of the $E(B-V)$ reddening from the \citet{SFD98} dust map (which is essentially the upper limit on reddening at high galactic latitudes), we adopt effective reddening $E(B-V)(1-\exp(-h/h_{\rm dust}))$, where $h$ is the distance from the Galactic plane, and $h_{\rm dust} = 130$ pc is the scale height of the dust layer \citep{gp98,kle11}. Such a small change is expected given the small values of extinction coefficients in near-IR W1 and W2 bands (0.13 and 0.17), the reddening distribution (0.18 mag in the 95th percentile), and the spatial distribution of RR Lyrae stars under consideration (most of them are located beyond a few scale heights of the galactic plane).

Briefly, the stars in our sample have metallicities ranging from $-0.1$ dex to $-2.6$ dex, and periods ranging from 0.36 to 0.73 days. The uncertainty in ${\rm [Fe/H]}$ is the same for all stars, $\sigma_{\rm [Fe/H]}=0.15$ dex (see Note 8 in Table 1 of \citealt{fer98}). Based on the analysis of \citet[][see their Appendix A]{kle11}, we assume that the uncertainty in log-period is $\sigma_{\rm \log P}=0.02\log P$. The uncertainty in $E(B-V)$ is assumed to be $\sigma_{\rm EBV} = 0.1E(B-V)$ \citep{SFD98}. The average uncertainty in apparent magnitudes is $\sigma_{\rm m} = 0.005$ mag \citep{kle14}.

The TGAS parallax measurements have reported uncertainties ranging from 0.22 mas to 0.47 mas (5th to 95th percentile), with a median of 0.28 mas. The median fractional uncertainty of our sample is $\sigma_{\rm \varpi}/\varpi=0.17$. As described by \citet{gdr2}, the reported parallax uncertainties were calculated as (their Equation 4)
\begin{equation}
\sigma_{\rm \varpi} = \sqrt{(f_{\rm \varpi}\varsigma_{\rm \varpi})^2 + \sigma_{\rm \varpi,add}^2}\label{TGAS_uncertainties},
\end{equation}
where $\varsigma_{\rm \varpi}$ is the formal parallax uncertainty, $f_{\rm \varpi}=1.4$, and $\sigma_{\rm \varpi,add}=0.2$ mas. To facilitate validation of TGAS parallax uncertainties in \autoref{Sec:Method}, we calculate $\varsigma_{\rm \varpi}$ values as
\begin{equation}
\varsigma_{\rm \varpi} = \sqrt{\sigma_{\rm \varpi}^2 - (0.20/1000)^2}/1.4\label{reported_TGAS_par_unc}
\end{equation}

The \citet{kle14} sample also contains 16 type $c$ RR Lyrae stars (i.e., first overtone pulsators; RRc) with measured W1 and W2 apparent magnitudes. However, we do not use these stars to constrain the PLZ relations of RRc stars because the sample is quite small, and because the median fractional parallax uncertainty of the sample is quite large ($\sigma_{\rm \varpi}/\varpi\approx0.5$). Instead, we use these stars to verify whether RRab and RRc stars follow the same PLZ relations. Before making such comparisons, the practice has been to first ``fundamentalize'' the periods of RRc stars (e.g., \citealt{dor04}). We follow the same practice, and calculate fundamentalized periods of RRc stars as
\begin{equation}
\log P = \log P_{fo} + 0.127\label{fund_P},
\end{equation}
where $P_{fo}$ is the original period.

The data used in this work are provided in a machine-readable format in the electronic version of the Journal (filename ``data.csv'').

\section{Method}\label{Sec:Method}

Using the above data set $\mathcal{D}$, we now wish to constrain parameters $a$, $b$, $M_{\rm ref}$, and $\sigma_{\rm M}$ that define the (noisy) PLZ relation
\begin{equation}
M = a\log(P/P_{\rm ref}) + b({\rm [Fe/H]} - {\rm [Fe/H]_{\rm ref}}) + M_{\rm ref} + \epsilon,\label{PLZ}
\end{equation}
where $M_{\rm ref}$ is the absolute magnitude at some reference period $P_{\rm ref}$ and metallicity ${\rm [Fe/H]_{\rm ref}}$ (here set to the median period and metallicity of the sample described in \autoref{Sec:Data}, $P_{\rm ref}=0.52854$ days and ${\rm [Fe/H]_{\rm ref}}=-1.4$ dex), and $a$ and $b$ scale the absolute magnitude with log-period and metallicity, respectively.

The $\epsilon$ is a standard normal random variable with zero mean and variance $\left(a\sigma_{\rm logP}\right)^2 + \left(b\sigma_{\rm [Fe/H]}\right)^2 + \sigma_{\rm M}^2$, where $\sigma_{\rm M}$ accounts for the scatter in absolute magnitude $M$ due to modeling uncertainties. When interpreting this scatter, however, it is important to keep in mind that, in reality, $\sigma_{\rm M}$ also {\em includes} unaccounted measurement uncertainties (added in quadrature). Thus, $\sigma_{\rm M}$ represents the so-called ``intrinsic'' scatter in a PLZ relation only if the measurement uncertainties are correctly estimated (which is difficult to do in practice).

As mentioned above, we also allow freedom in the error model used for the parallax measurements. We model the TGAS parallax measurements as being drawn from a Gaussian distribution centered on
\begin{equation}
\varpi^\prime = 1/r + \varpi_{\rm 0}\label{model_varpi},
\end{equation}
where $r$ is the true heliocentric distance (in parsecs), and $\varpi_{\rm 0}$ represents the global offset of TGAS parallaxes with respect to the inverse distances (e.g., an offset due to the impact of imperfectly modeled basic-angle variations on the astrometric solution; \citealt{gdr2}). The standard deviation of this Gaussian is equal to the uncertainty in the TGAS parallax, which we model using \autoref{TGAS_uncertainties}, where $f_{\rm \varpi}$ and $\sigma_{\rm \varpi,add}$ are also included as free parameters. The expressions defined by Equations~\ref{model_varpi} and~\ref{TGAS_uncertainties} were motivated by conclusions of recent studies that have found TGAS parallaxes to be offset \citep{jao16,st16,Dri16}, and studies that have found TGAS parallax uncertainties to be overestimated \citep{gks16,cas16}.\footnote{Note that these models can be easily extended: For example, we could model the dependence of parallax on ecliptic latitude by simply adding a $b_{\rm \varpi}\beta$ term to \autoref{model_varpi}, where $b_{\rm \varpi}$ is a new model parameter, and $\beta$ is the ecliptic latitude of a star (in units of arcsec).}

To constrain the PLZ relation in a probabilistic manner we need to calculate the joint posterior probability $p(\plzpars, \tgaspars, L, \{\starpars\} \given \mathcal{D})$, given the data set $\mathcal{D}$, of the PLZ parameter value set $\plzpars = \{a, b, M_{\rm ref}, \sigma_{\rm M} \}$, the TGAS parallax validation parameters $\tgaspars = \{\varpi_{\rm 0}, f_{\rm \varpi}, \sigma_{\rm \varpi,add}\}$, the scale length parameter $L$ (used in the distance prior, see \autoref{distance_prior} below), and the set of nuisance parameters $\starpars = \{r, \log P^{\rm int}, {\rm [Fe/H]^{\rm int}}, EBV^{\rm int}\}_k$ that represent the true distance $r$, intrinsic log-period $\log P^{\rm int}$, metallicity ${\rm [Fe/H]^{\rm int}}$, and reddening $EBV^{\rm int}$ for each star. For conciseness, we also define $\bs{\theta} = (\plzpars, \tgaspars, L)$. For this work, our main interest is in the marginal posterior probability of the TGAS validation and PLZ parameters $p(\bs{\theta} \given \mathcal{D})$, which is related to the marginal likelihood $p(\mathcal{D} \given \bs{\theta})$ through
\begin{equation}
p(\bs{\theta} \given \mathcal{D}) \propto p(\mathcal{D} \given \bs{\theta}) \, p(\bs{\theta})\label{posterior},
\end{equation}
where $p(\bs{\theta})$ is the prior probability of the parameter value set $\bs{\theta}$, and
\begin{eqnarray}
p(\mathcal{D} \given \bs{\theta}) &=& \prod_k p({\bf d_k} \given \bs{\theta})\label{likelihood}
\\
&=& \prod_k \, \int p({\bf d_k} \given \bs{\theta}, \bs{\alpha}_k) \, p(\bs{\alpha}_k) \, \dd \bs{\alpha}_k
\end{eqnarray}
is the marginal likelihood of the full data set given the combined PLZ, TGAS validation, and $L$ parameters ($\bs{\theta}$), and assuming independent data points.

The (un-marginalized) likelihood function for the $k$th star is given as follows:
\begin{equation}
\begin{split}
p&({\bf d_k} \given \bs{\theta}, \bs{\alpha}_k) = p\left(\varpi \given r, \tgaspars\right) \, p\left(\log P \given \log P^{\rm int}\right)  \\
&\times p\left(EBV \given EBV^{\rm int}\right) \, p\left({\rm [Fe/H]} \given {\rm [Fe/H]}^{\rm int}\right) \\
&\times p(m \given A_\lambda, \starpars, \plzpars),
\label{one_star_likelihood}
\end{split}
\end{equation}
where:
\begin{equation}
\mathcal{N}(x \given \mu, \sigma^2) = \frac{1}{\sqrt{2\pi\sigma^2}} \, \exp\left(-\frac{1}{2}\frac{(x-\mu)^2}{\sigma^2}\right)
\end{equation}
is a normal distribution centered on $\mu$ with variance $\sigma^2$, and
\begin{align}
&p\left(\varpi \given r, \tgaspars\right) = \mathcal{N}\left(\varpi \given \varpi^\prime, \sigma^2_{\rm \varpi}\right)\\
&p\left(\log P \given \log P^{\rm int}\right) = \mathcal{N}\left(\log P \given \log P^{\rm int}, \sigma^2_{\rm \log P}\right)\label{P_int}  \\
&p\left({\rm [Fe/H]} \given {\rm [Fe/H]}^{\rm int}\right) = \mathcal{N}\left({\rm [Fe/H]} \given {\rm [Fe/H]}^{\rm int}, \sigma^2_{\rm [Fe/H]}\right) \\
&p\left(EBV \given EBV^{\rm int}\right) = \mathcal{N}\left(EBV \given EBV^{\rm int}, \sigma^2_{\rm EBV}\right)\label{EBV_int} \\
&p\left(m \given A_\lambda, \starpars, \plzpars\right) = \mathcal{N}\left(m \given m^\prime, \sigma^2_m+\sigma^2_{\rm M}\right) \\
&m^\prime = M^{\rm int} + A_\lambda EBV^{\rm int} + 5\log r - 5 \label{m_int} \\
&M^{\rm int} = a\log\left(P^{\rm int}/P_{\rm ref}\right) + b\left({\rm [Fe/H]}^{\rm int} - {\rm [Fe/H]_{\rm ref}}\right) + M_{\rm ref}
\end{align}

\begin{figure}
\plotone{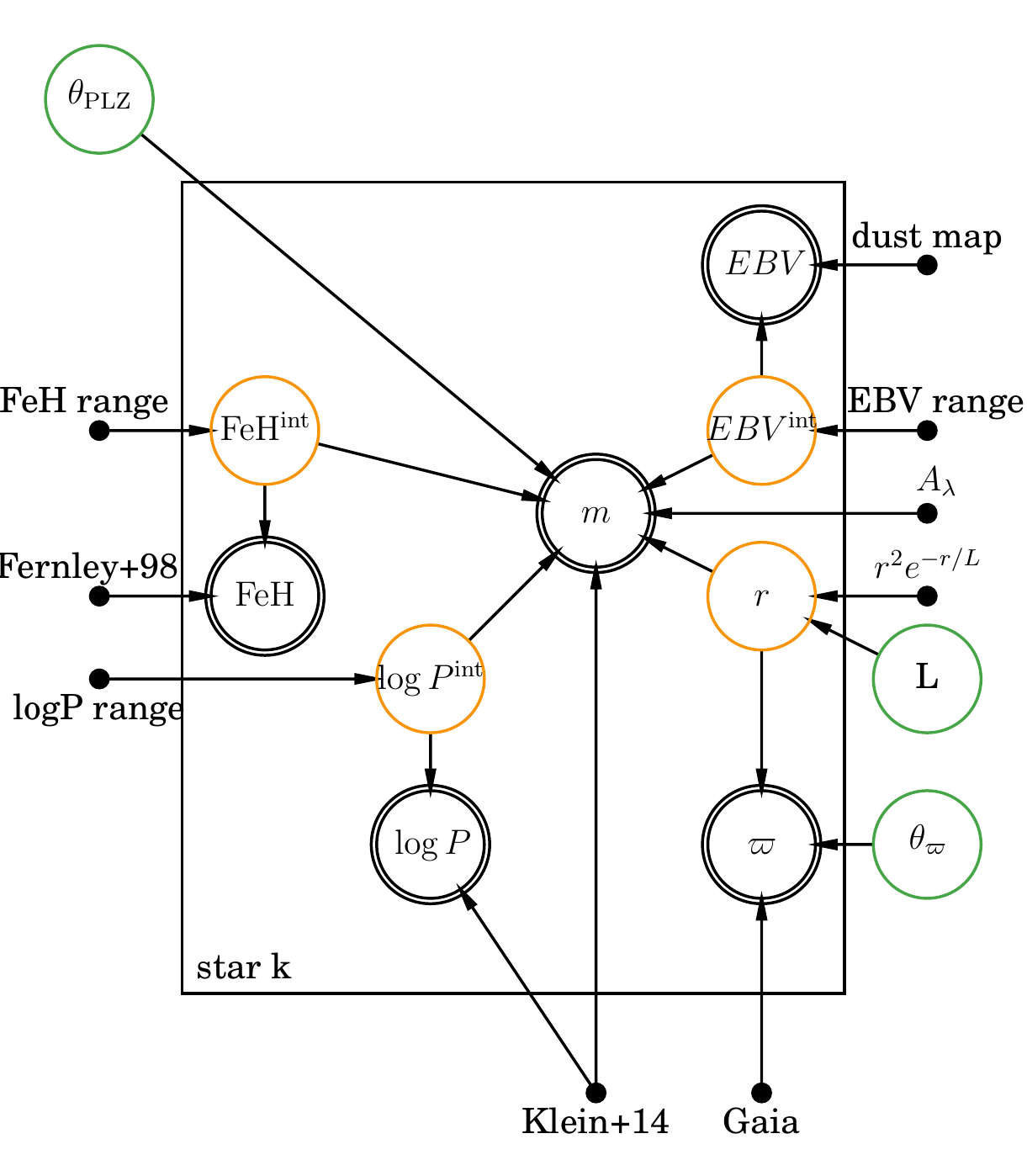}
\caption{
The probabilistic graphical model that describes the dependencies between model parameters and data used in this work. Double circles indicate likelihoods, single orange circles indicate nuisance parameters, while single green circles indicate model parameter sets. Fixed parameters, such as the extinction coefficient \citep[$A_\lambda$;][]{sch16}, priors on nuisance parameters (e.g., $\log P^{\rm int}$ range), as well as data sources (e.g., Gaia), are not enclosed in circles. Parameters inside the square are specific to the $k$th star, while those on the outside are global. The arrows indicate conditional dependence. For example, the arrows from $r$ and $\theta_{\rm \varpi}$ to $\varpi$ indicate that the observed parallax $\varpi$ depends on the heliocentric distance $r$ and TGAS parallax validation parameters $\theta_{\rm \varpi}$ (i.e., $p(\varpi \given r, \tgaspars)$).
\label{PLZ_PGM}}
\end{figure}

Our statistical model is also visualized as a probabilistic graphical model (PGM) shown in \autoref{PLZ_PGM}. One of the advantages of PGMs is that they enable a straightforward examination of dependencies between data and model parameters \citep{pgm_book}. For example, while the observed apparent magnitude $m$ depends on the adopted extinction coefficient $A_\lambda$, and parameters in $\starpars$ and $\plzpars$ sets, the observed parallax $\varpi$ depends only on distance and TGAS parallax validation parameters, $\theta_{\rm \varpi}$, described above.

In Equations~\ref{P_int} to~\ref{EBV_int}, we model the observed log-period, metallicity, and reddening as being drawn from Gaussian distributions centered on some ``intrinsic'' values (represented by the superscript ``int''), and with a standard deviation equal to the uncertainty in measurement or the model. The ``int'' parameters are ``intrinsic'' in the sense that they are not affected by uncertainties (model or observational).

Note that \autoref{one_star_likelihood} contains 7 global parameters (4 in $\plzpars$ and 3 in $\tgaspars$ parameter sets), and 4 {\em nuisance} parameters for each star, $\starpars$. Since we are (at the moment) not interested in these nuisance parameters, we marginalize (i.e. integrate) \autoref{one_star_likelihood} over these parameters.

We use the following prior probability distributions for the nuisance parameters $\starpars$. For ${\rm [Fe/H]^{\rm int}}$ and $\log P^{\rm int}$ we choose priors that are uniform in ranges appropriate for RR Lyrae stars: 0 dex to -3 dex, and -1.0 $\log ({\rm day})$ to 0 $\log ({\rm day})$. For $EBV^{\rm int}$, we adopt a uniform prior in the 0 mag to 1 mag range (appropriate for low-extinction regions, as is the case here), and for distance we adopt an exponentially decreasing volume density prior with a scale length parameter $L$ \citep{b-j15,abj16}
\begin{equation}
p(r \given L) = 1/(2L^3) \, r^2 \, \exp(-r/L)\label{distance_prior}.
\end{equation}
The prior $p(r \given L)$ is positive in the $200 < r/{\rm pc} < 2700$ distance range, and zero elsewhere.

The marginal likelihood for the $k$th star is
\begin{equation}
p\left({\bf d_k} \given \bs{\theta}\right) = \int \, p\left({\bf d_k} \given \bs{\theta}, \bs{\alpha}_k\right) \, p\left(\bs{\alpha}_k\right) \, \dd \bs{\alpha}_k \label{marginal_likelihood1}
\end{equation}
which can be written in a simpler form by analytically performing the integrals over the Gaussian expressions in $\log P^{\rm int}$, ${\rm [Fe/H]^{\rm int}}$, and $EBV^{\rm int}$:
\begin{eqnarray}
p\left({\bf d_k} \given \bs{\theta}\right) &=& \, \int^{2700\, pc}_{200\, pc} \, \dd r \, \mathcal{N}\left(\varpi \given \varpi^\prime, \sigma^2_{\rm \varpi}\right) \nonumber \\ 
&& \times \mathcal{N}\left(DM^\prime \given DM, \sigma^2_{\rm DM}\right) \, p(r \given L)\label{marginal_likelihood2} ,
\end{eqnarray}
where
\begin{align}
DM &= 5\log r - 5 \nonumber \\
DM^\prime &= m - A_\lambda EBV - \nonumber \\
          &\left(a\log(P/P_{\rm ref}\right) + b\left({\rm [Fe/H] - [Fe/H]_{ref}}\right) + M_{\rm ref}) \nonumber \\
\sigma_{\rm DM}^2 &= \sigma_m^2 + \left(a\sigma_{\rm logP}\right)^2 + \left(b\sigma_{\rm [Fe/H]}\right)^2 + \left(A_\lambda \sigma_{\rm EBV}\right)^2 + \sigma_{\rm M}^2\label{sigma_DM}
\end{align}
The likelihood for the entire data set $\mathcal{D}$ can now be calculated using \autoref{likelihood}.

Before we can calculate the (marginal) posterior distribution (\autoref{posterior}), we need to define prior probabilities for the global parameters $\plzpars$, $\tgaspars$, and $L$. For $\sigma_{\rm M}$, $\sigma_{\rm \varpi, add}$, and $L$ parameters, we adopt Jeffreys log-uniform priors \citep[$p(x)\propto1/x$; ][]{jay68}, and for the $b$ parameter that scales the absolute magnitude with ${\rm [Fe/H]}$, we adopt a uniform prior that is positive for $0 < b/{\rm mag\, dex^{-1}} < 3$ (based on stellar evolution and pulsation models of RR Lyrae stars; \citealt{mar15}). Since the $\varpi^\prime$ parameter must be positive in the 200 pc to 2700 pc range (\autoref{model_varpi}), for $\varpi_{\rm 0}$ we adopt a uniform prior that is positive in the $-5 < \varpi_{\rm 0}/{\rm mas} < 20$ range. For the remaining model parameters we adopt wide uniform priors.

To efficiently explore the parameter space, we use the \citet{gw10} Affine Invariant Markov chain Monte Carlo (MCMC) Ensemble sampler as implemented in the \texttt{emcee} package\footnote{\url{http://dan.iel.fm/emcee/current/}} (v2.2.1, \citealt{fm13}). We use 160 walkers and obtain convergence\footnote{We checked for convergence of chains by examining the auto-correlation time of the chains per dimension.} after a burn-in phase of 1000 steps per walker. The chains are then evolved for another 1500 steps, and the first 1000 (burn-in) steps are discarded.

To describe the marginal posterior distributions of individual model parameters, we measure the median, the difference between the 84th percentile and the median, and the difference between the median and the 16th percentile of each marginal posterior distribution (for a Gaussian distribution, these differences are equal to $\pm1$ standard deviation). We report these and maximum a posteriori (MAP) values in \autoref{PS1_PLZ}.

\capstartfalse
\begin{deluxetable*}{lccccccc}
\tablecolumns{8}
\tablecaption{PLZ Relations for RRab Stars and TGAS Parallax Validation Parameters\label{PS1_PLZ}}
\tablehead{
\colhead{Band} & \colhead{a} & \colhead{b} & \colhead{$M_{\rm ref}$} & \colhead{$\sigma_{\rm M}$} & \colhead{$\varpi_{\rm 0}$} & \colhead{$f_{\varpi}$} & \colhead{$\sigma_{\rm \varpi,add}$} \\
\colhead{ } & \colhead{(mag dex$^{-1}$) } & \colhead{(mag dex$^{-1}$)} & \colhead{(mag)} & \colhead{(mag)} & \colhead{(mas)} & \colhead{ } & \colhead{(mas)}
}
\startdata
W1 & $-2.47^{+0.74}_{-0.73}$ & $0.15^{+0.09}_{-0.08}$ & $-0.42^{+0.12}_{-0.10}$ & $0.07^{+0.08}_{-0.05}$ & $-0.05^{+0.05}_{-0.06}$ & $0.86^{+0.34}_{-0.31}$ & $0.16^{+0.04}_{-0.05}$ \\
 & $-2.28$ & 0.22 & $-0.46$ & 0.04 & $-0.04$ & 0.81 & 0.16 \\
W2 & $-2.40^{+0.84}_{-0.82}$ & $0.17^{+0.10}_{-0.09}$ & $-0.52^{+0.11}_{-0.10}$ & $0.05^{+0.07}_{-0.03}$ & $0.00^{+0.05}_{-0.05}$ & $0.73^{+0.33}_{-0.34}$ & $0.18^{+0.03}_{-0.04}$ \\
 & $-2.15$ & 0.24 & $-0.59$ & 0.05 & $0.02$ & 0.75 & 0.18 \\
\enddata
\tablecomments{In each band, the first line lists the median, the difference between the 84th percentile and the median, and the difference between the median and the 16th percentile of each marginal posterior distribution. The second line provides the maximum a posteriori (MAP) values.} 
\end{deluxetable*}
\capstarttrue

\section{Results}\label{Sec:Results}

\begin{figure*}
\plotone{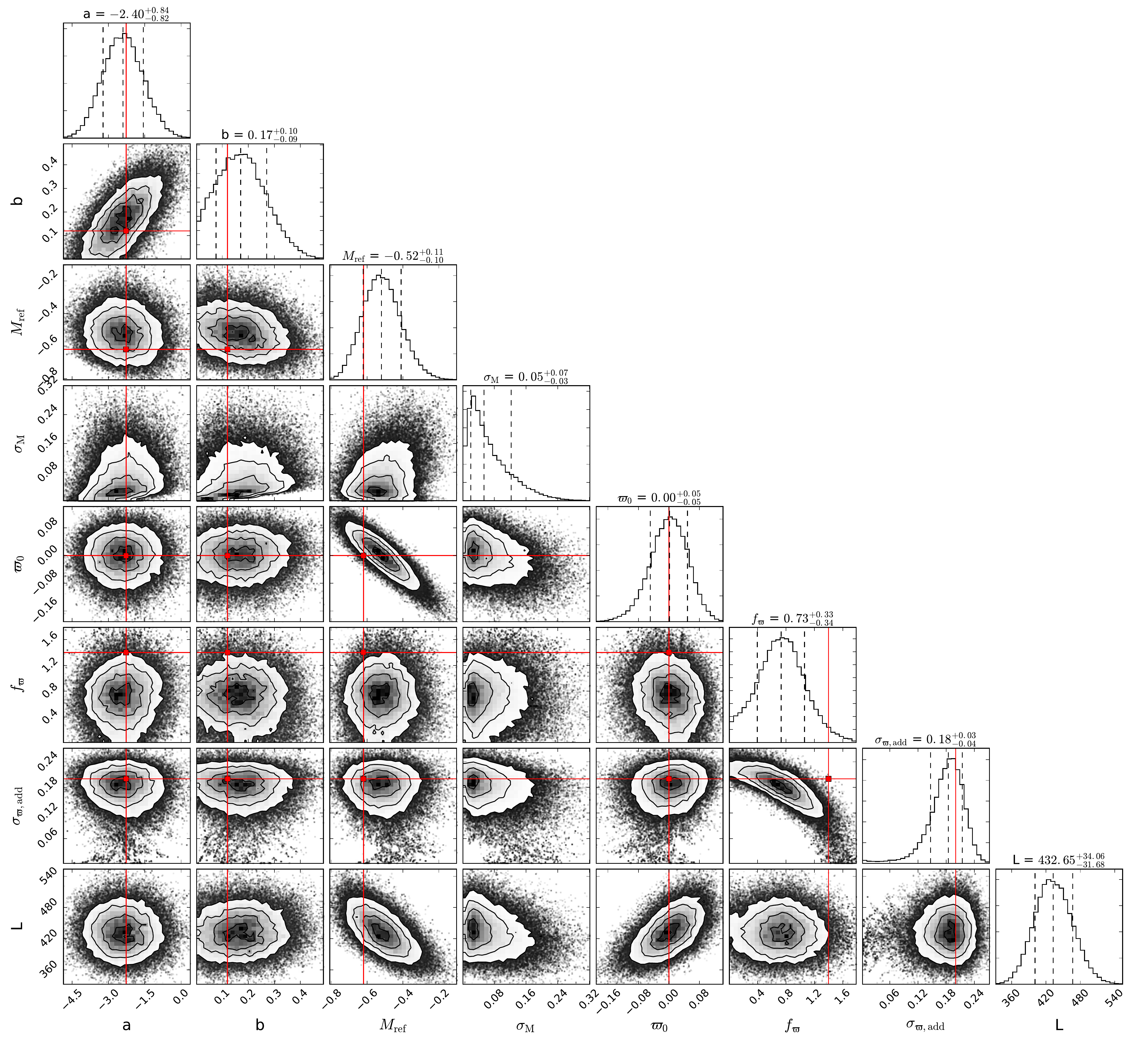}
\caption{
Posterior distributions of parameters in $\theta_{\rm PLZ}$ and $\theta_{\rm \varpi}$ sets, for the parameters constrained using W2 band data. The histograms show the marginalized posterior distributions for each parameter, with the dashed lines indicating the 16th percentile, the median, and the 84th percentile. The contour plots show the two-dimensional posterior distributions for parameter pairs, with the contours indicating $0.5\sigma$, $1\sigma$, $1.5\sigma$, and $2\sigma$ levels. Uncorrelated pairs have roundish contours, while correlated parameters like $f_{\rm \varpi}$ vs.~$\sigma_{\rm \varpi,add}$ have elongated distributions. For comparison, the red squares and lines show the best-fit values of $a$, $b$, and $M_{\rm ref}$ measured by \citet{dam14}, the $\varpi_{\rm 0}=0$ mas line, and the $f_{\rm \varpi}$ and $\sigma_{\rm \varpi, add}$ values adopted by \citet{gdr2}.
\label{corner_plot}}
\end{figure*}

In this Section, we discuss PLZ relations and TGAS parallax validation parameters that we have constrained by applying the method described in \autoref{Sec:Method}, to data described in \autoref{Sec:Data}. To illustrate the correlations between various model parameters, in \autoref{corner_plot} we show two-dimensional posterior distributions for parameter pairs. The parameters we constrain also allow us to calculate more precise distances to RR Lyrae stars in our sample (\autoref{table_distances}).

\subsection{PLZ Parameters}

We find that, within the uncertainties, the PLZ parameters obtained using W1 band data are consistent with those obtained using W2 band data. This result is not too surprising given i) fairly large uncertainties in parameters, ii) the proximity of W1 and W2 bands in wavelength, and iii) the fact that the parameters of PL(Z) relations change little with wavelength for bands redder than the $H$ or $K$ band (\citealt{cat04,mar15}; Figure 4 of \citealt{mad13}). Since the two PLZ relations are consistent, we use PLZ parameters for the W2 band when comparing relations to previous studies.

Overall, our PLZ relations are consistent with PL(Z) relations found by previous studies (within $1\sigma$ of uncertainties). For example, \citet{mad13}, \citet{kle14}, and \citet{dam14} measure the period slope of the W2 band PLZ relation for RRab stars to be $a=-2.6\pm0.9$, $a=-2.4\pm0.2$, and $-2.3\pm0.1$, respectively, while we find $a=-2.4\pm0.8$. Our estimate of the absolute magnitude at $P = 1$ day of $-1.2\pm0.2$ mag, is consistent with $\approx-1.1\pm0.1$ mag estimated by the above studies. Regarding the metallicity slope of the PLZ relation, our measurement of $b=0.17\pm0.10$ is consistent at the $1\sigma$ level with the \citet{dam14} value of $0.12\pm0.02$, and (in)consistent at the $2\sigma$ level with the slope of zero reported by \citet{mad13} and \citet{kle14} (note: the latter two studies did not report the uncertainty of their estimate).

The $\sigma_{\rm M}$ MAP values\footnote{Since the marginal posterior distribution of $\sigma_{\rm M}$ is quite asymmetric (\autoref{corner_plot}), we discuss the MAP value as it is closer to the mode of the marginal distribution.} of 0.04 mag and 0.05 mag for W1 and W2 bands, respectively, seem a bit high given the expectation of a small intrinsic scatter in near-IR PLZ relations (Table 6 of \citealt{mar15}). However, recall that $\sigma_{\rm M}$ {\em includes} (via \autoref{sigma_DM}) unaccounted measurement uncertainties in apparent magnitude, metallicity, log-period, and reddening. One possible source of unaccounted uncertainties may be the scatter in the WISE magnitude zeropoint, which is $\approx0.03$ mag (Table 8 of Section 4.4.h Explanatory Supplement to the WISE All-Sky Data Release Products\footnote{\url{http://wise2.ipac.caltech.edu/docs/release/allsky/expsup/sec4_4h.html}}). Another source may be the uncertainty in ${\rm [Fe/H]}$, which \citet{fer98} estimated at 0.15 dex, but that could be higher since the ${\rm [Fe/H]}$ values reported by \citet{fer98} are actually a compilation of metallicities measured by different studies.

\subsection{TGAS Parallax Validation Parameters}

We measure the $\varpi_{\rm 0}$ parameter (that models the global offset of TGAS parallax measurements) to be consistent with zero within 0.05 mas (i.e., $\varpi_{\rm 0} = 0.00\pm0.05$ mas), indicating that there is no statistically significant offset in parallax, at least for this sample of distant RR Lyrae stars (median distance $\approx1$ kpc).

Five RR Lyrae stars in our sample (SU Dra, RR Lyr, UV Oct, XZ Cyg, and RZ Cep) also have parallaxes measured by \citet{ben11} using Hubble Space Telescope (HST) astrometric observations. We find the average of the difference between TGAS and HST parallax measurements to be 0.04 mas, in agreement with our $\varpi_{\rm 0} = 0.00\pm0.05$ mas measurement.

Regarding the renormalization of TGAS formal parallax uncertainties $\varsigma_{\rm \varpi}$ (see \autoref{TGAS_uncertainties}), we find $\sigma_{\rm \varpi,add} = 0.18$ mas and that $f_{\rm \varpi}$ is consistent with 1 within uncertainties(i.e., $f_{\rm \varpi}=0.7\pm0.3$). In principle, it is possible for $f_{\rm \varpi}$ to be less than 1, if the {\em formal} TGAS parallax uncertainties, $\varsigma_{\rm \varpi}$, are overestimated. Since $\varsigma_{\rm \varpi}$ are calculated from the inverse $5\times5$ normal matrix of the astrometric parameters \citep{gdr2}, this could happen if there is some serious problem with the software that performs this calculation. As we do not believe this to be likely, we assume that $f_{\rm \varpi}\geq1$, or given data, $f_{\rm \varpi} = 1$.

\subsection{Do RRc Stars Follow the PLZ Relation for RRab Stars?}\label{RRc_validation}

Since we did not use RRc stars to constrain the model described in \autoref{Sec:Method} (only RRab stars were used), we can now use this model to verify whether RRc stars follow the same PLZ relation as RRab stars.

Given our data and the model, the most appropriate comparison between the data and the model is the one between the observed TGAS parallaxes, $\varpi$, and inverse distances, $1/r$. To infer the inverse distance, we first need to calculate the marginal posterior distribution of the heliocentric distance for the $k$th star, $p( r_k \given {\bf d_k}, \bs{\theta})$, since the marginal posterior distribution of the inverse distance, $p( 1/r_k \given {\bf d_k}, \bs{\theta})$, is equal to
\begin{equation}
p( 1/r_k \given {\bf d_k}, \bs{\theta}) = r_k^2 \, p( r_k \given {\bf d_k}, \bs{\theta})\label{varpi_posterior}.
\end{equation}

Using relations of conditional probability, the marginal posterior distribution of the heliocentric distance for the $k$th star can be written as
\begin{eqnarray}
p( r_k \given {\bf d_k}, \bs{\theta}) &\propto& \int \, \dd \bs{\theta} \, \mathcal{N}(\varpi \given \varpi^\prime, \sigma^2_{\rm \varpi}) \nonumber \\
&& \times \mathcal{N}(DM^\prime \given DM, \sigma^2_{\rm DM}) \, p(r \given L)\label{distance_posterior},
\end{eqnarray}
where \autoref{distance_posterior} is being integrated (i.e., marginalized) over the scale length parameter $L$, as well as parameters contained in $\plzpars$ and $\tgaspars$ sets\footnote{The marginalization over $\log P^{\rm int}$, ${\rm [Fe/H]}^{\rm int}$, and $EBV^{\rm int}$ parameters is implicit in the expression for $\mathcal{N}(DM^\prime \given DM, \sigma^2_{\rm DM})$ (see \autoref{marginal_likelihood1}).}.

We evaluate \autoref{varpi_posterior} over the $1/(2700\, {\rm pc}) < (1/r) / {\rm arcsec} < 1/(200\, {\rm pc})$ inverse distance range and average it over 160 samples of $L$, $\plzpars$, and $\tgaspars$ parameters taken from the last step of the Markov chain (which has 160 walkers). The mean inverse distance and its uncertainty are obtained by fitting a Gaussian to $p( 1/r_k \given {\bf d_k}, \bs{\theta})$ evaluated on the above grid. Similarly, we evaluate \autoref{distance_posterior} over the $200 < r /{\rm pc} < 2700$ distance range, and list the distances and their uncertainties in \autoref{table_distances}. The median fractional uncertainty in distance is 6\%.

\capstartfalse
\begin{deluxetable}{lc}
\tablecolumns{2}
\tablecaption{Distances to RR Lyrae Stars\label{table_distances}}
\tablehead{
\colhead{Name} & \colhead{Heliocentric distance} \\
\colhead{ } & \colhead{(pc)}
}
\startdata
RRLyr & $265.8\pm10.7$ \\
FWLup & $369.4\pm19.1$ \\
CSEri* & $475.3\pm29.8$
\enddata
\tablecomments{Distances to RRc stars (stars with ``*'' in their names) may be biased (see \autoref{RRc_validation}). A machine readable version of this table is available in the electronic edition of the Journal. A portion is shown here for guidance regarding its form and content.} 
\end{deluxetable}
\capstarttrue

\begin{figure}
\plotone{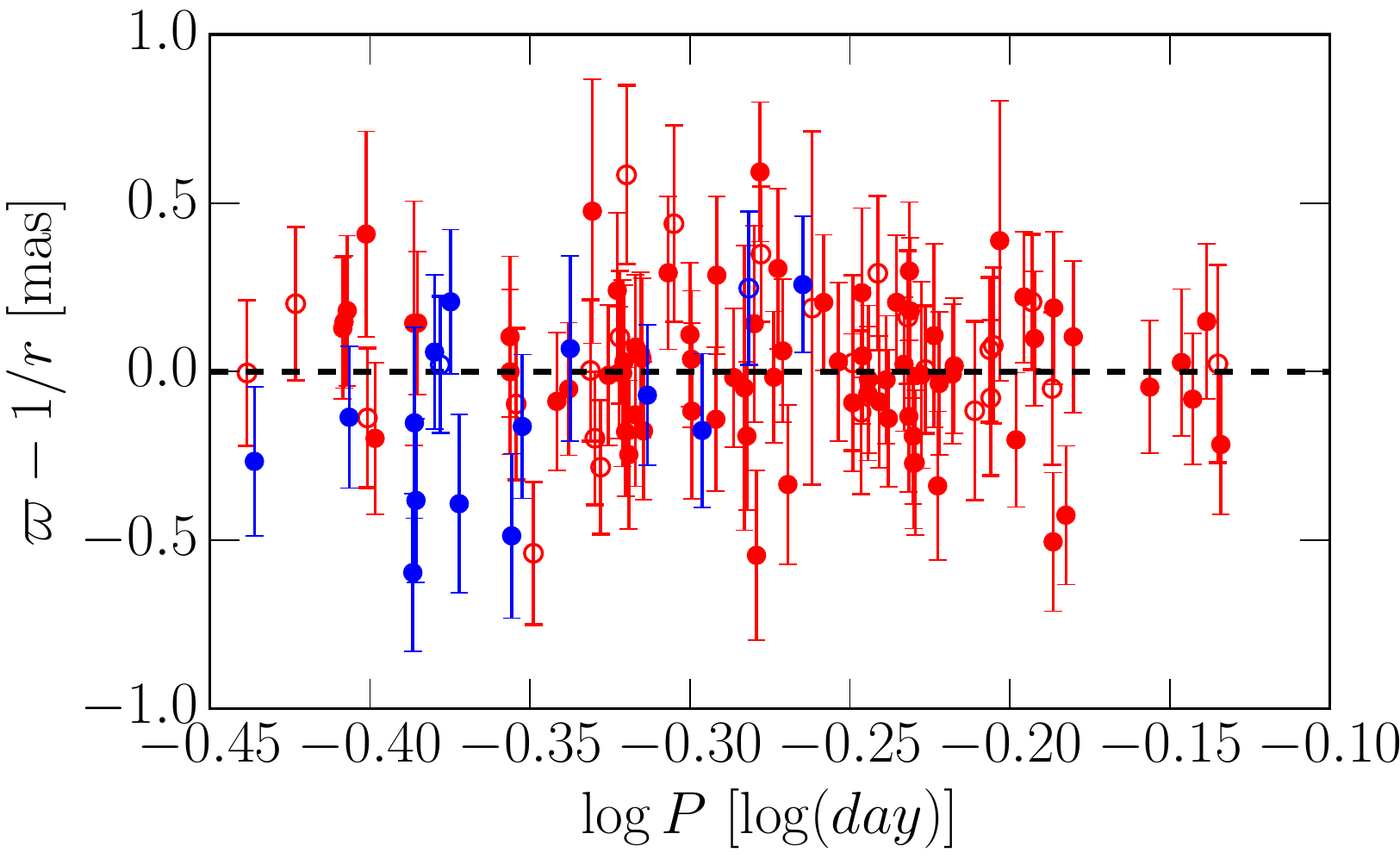}
\plotone{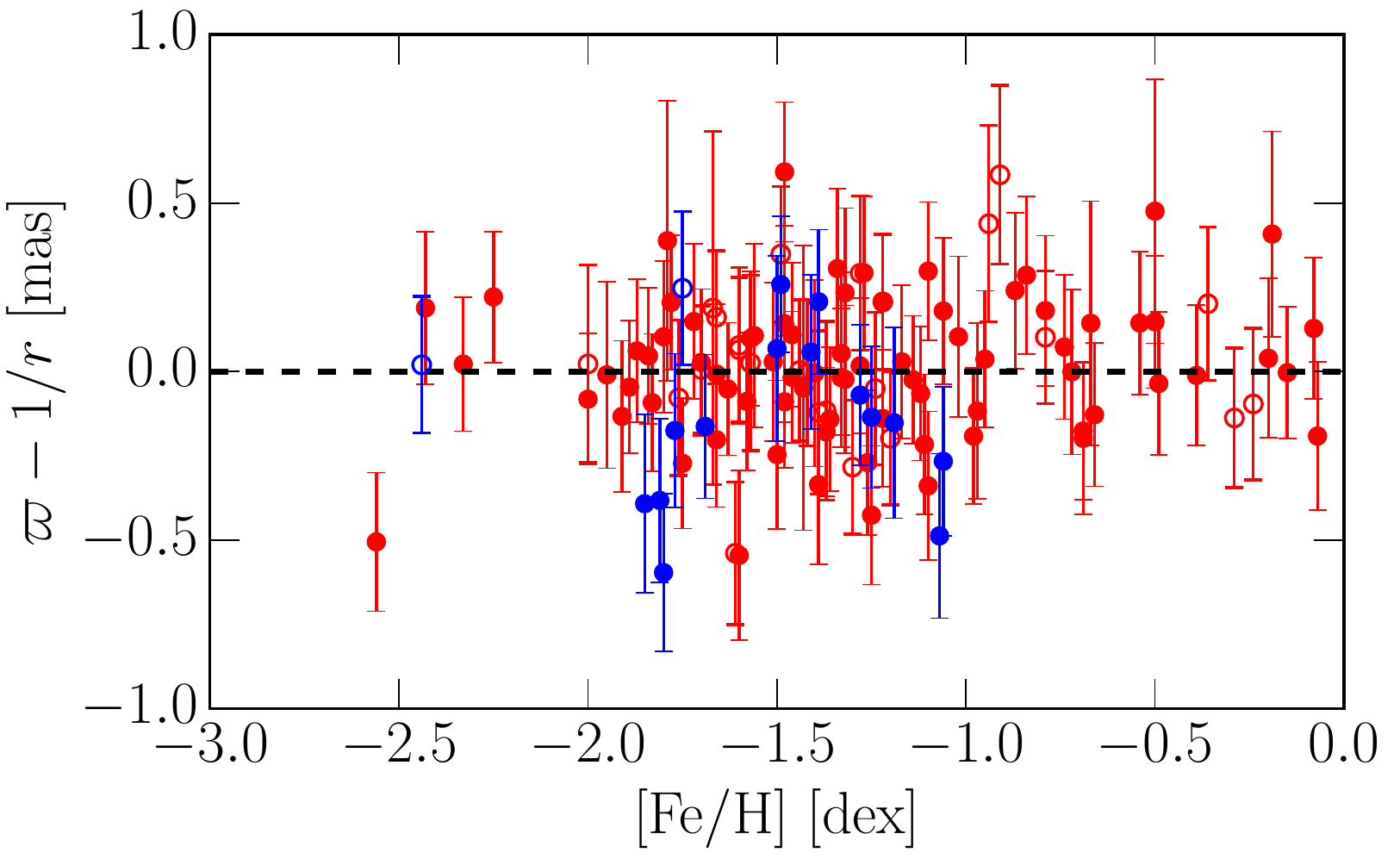}
\caption{
The symbols with error bars illustrate the difference between the observed TGAS parallax ($\varpi$) and the inverse distance inferred from our model (which was constrained using RRab stars), as a function of $\log P$ ({\em top panel}) and ${\rm [Fe/H]}$ ({\em bottom panel}). The errobars indicate the quadratic sum of the uncertainty in the observed parallax and the inferred inverse distance. The RRab and RRc stars are shown as red and blue symbols, respectively. The RR Lyrae stars whose light curves are affected by the \citet{bla07} effect (i.e., the amplitude or phase modulation of the light curve; as identified via \url{http://www.univie.ac.at/tops/blazhko/Blazhkolist.html}) are denoted with open circles, and the stars not known to exhibit the Bla\v{z}ko effect are denoted with solid circles. For RRc stars, we are showing fundamentalized periods (\autoref{fund_P}). Note how the observed parallaxes of RRc stars seem to systematically deviate from the inferred inverse distances, suggesting that RRc stars may not follow the same PLZ relations as RRab stars. Also, note that Bla\v{z}ko-affected stars do not seem to scatter more than stars not known to exhibit the Bla\v{z}ko effect, indicating that the inclusion of Bla\v{z}ko-affected stars does not significantly increase the scatter in the PLZ relations (i.e., the $\sigma_{\rm M}$ parameter).
\label{validation_plots}}
\end{figure}

As the top panel in \autoref{validation_plots} shows, the inferred inverse distances of RRc stars seem to systematically deviate from observed TGAS parallaxes as the $\log P$ decreases (i.e., as the periods get shorter). This trend suggests that the PLZ relation of RRc stars may have a shallower slope than the PLZ relation of RRab stars. Indeed, when both RRab and RRc stars are used to constrain the PLZ relations in the W1 and W2 bands (periods of RRc stars are fundamentalized), the resulting period slopes are shallower, though still consistent (within $1\sigma$) with the period slopes obtained using only RRab stars ($a_{\rm RRab+RRc}=-1.5\pm0.6$ mag dex$^{-1}$ vs.~$a_{\rm RRab}=-2.4\pm0.8$ mag dex$^{-1}$). 

The above result (i.e., the flattening of the period slope when RRc stars are added) is consistent with the finding of \citet{kle14} that PLZ relations of RRc stars have shallower period slopes in W1 and W2 bands than RRab stars (though, the RRab and RRc period slopes they measure are consistent with each other within $1\sigma$). On the other hand, the above result is at odds with \citet{bra15}, who find that the period slopes of RRab and RRc stars in the $K-$band\footnote{The period slope is not expected to change significantly between the $K$ and WISE bands (see Figure 4 of \citealt{mad13}).} are quite consistent, $a\approx-2.4\pm0.2$ mag dex$^{-1}$.

Relative to the observed TGAS parallaxes, the inverse distances of RRc stars also show an interesting V-shaped trend as a function of ${\rm [Fe/H]}$ (bottom panel of \autoref{validation_plots}). If real, this trend would suggest that, for RRc stars, the metallicity slope of the PLZ relation changes sign at ${\rm [Fe/H]}\approx-1.5$ dex.

In conclusion, while the behavior of RRc stars in \autoref{validation_plots} suggests that RRc stars may not follow the same PLZ relation as RRab stars, the evidence is not strong; the sample of RRc stars we use is quite small (only 16 stars) and their parallaxes are fairly uncertain. In the future, we intend to reexamine the PLZ relations of RRc stars using more precise Gaia DR2 parallaxes and a larger set of RRc stars (e.g., the one published by \citealt{gav14}).

\section{Summary and Conclusions}\label{Sec:Conclusions}

In this work, we have presented a probabilistic method that simultaneously constrains a period-luminosity-metallicity (PLZ) relation and validates (TGAS) parallax measurements. Compared to the traditionally-used weighted least-squares fitting of a PLZ relation, our approach allows for a direct usage of parallax (and other) measurements, while accounting fully for their precision. In comparison, the traditional approach cannot use imprecise parallaxes as their transformation into distance, and the subsequent characterization of the uncertainty in distance, is not a trivial task \citep{b-j15,abj16}.

The final product of our approach is the full posterior distribution of all model parameters (in the form of a Markov chain). The posterior distribution enables a more general description of model parameters and their correlations (see \autoref{corner_plot}). For example, the marginal posterior distribution (i.e., the histogram) of $\sigma_{\rm M}$ in \autoref{corner_plot} clearly shows that this parameter cannot be described by a Gaussian, which is something the traditional approach would be forced to do. Furthermore, the non-linear correlation between $f_{\rm \varpi}$ and $\sigma_{\rm \varpi,add}$ is beautifully described by the joint posterior distribution of these two parameters (see \autoref{corner_plot}). The least-squares approach would be hard-pressed to capture such complexities.

We have used this probabilistic approach to constrain the PLZ relations in the near-IR W1 and W2 bands used by the WISE mission. Overall, the PLZ parameters we recover are consistent (within uncertainties) with the parameters found by previous studies \citep{mad13,kle14,dam14}. Due to the fairly high fractional parallax uncertainty of our sample (median $\sigma_{\rm \varpi}/\varpi \approx 0.17$), we were not able to constrain the slopes in log-period and metallicity more precisely than \citet{dam14}. However, since \citet{dam14} used iterative least-squares fitting to constrain PLZ relations and fairly {\em ad hoc} removal of outlying data points, it is possible that their uncertainties may be underestimated. The second release of {\em Gaia} data, scheduled for April 2018, will provide more accurate and precise parallax measurements, and based on our current experience, is expected to place significantly tighter constraints on PLZ relations of RR Lyrae and other pulsating stars.

When fitting PLZ relations, we did not reject any outlying data points. Judging by the size of error bars and the distribution of $\varpi-1/r$ values of RRab stars in \autoref{validation_plots}, there does not seem to be many potential outliers that could have biased our results. However, as the precision of {\em Gaia} parallaxes improves, outliers may appear and unless they are properly handled, they may bias the measurement of PLZ relation parameters. Within the probabilistic framework presented in this work, inliers and outliers can be modeled using a {\em mixture} model (e.g., see Section 3 and Equation 17 of \citealt{hbl10}). As an example of how a mixture model can be used to model inliers and outliers in the context of constraining PLZ relations, we refer the interested reader to Appendix B of \citet{ses17}.

Since obtaining precise and accurate measurements of ${\rm [Fe/H]}$ for RR Lyrae stars is not a trivial task (e.g., see \citealt{nem13}), the optimal data set for constraining PLZ relations for RR Lyrae stars may need to contain field {\em and} globular cluster RR Lyrae stars. By being at the same distances and by having the same (and precisely measured) ${\rm [Fe/H]}$, globular cluster RR Lyrae stars could be used to constrain the period and metallicity dependence of the PLZ relation (i.e., parameters $a$ and $b$), while a few field RR Lyrae stars with well-measured Gaia parallaxes and ${\rm [Fe/H]}$ (e.g., those observed by \citealt{nem13}) would constrain the zero-point of the PLZ relation. Such datasets were used by \citet{sol06} and \citet{dam14}, but these studies used least-squares fitting to constrain PLZ relations. In the future, we may apply our method to the same datasets.

While constraining PLZ relations, we also simultaneously constrained parameters that model TGAS parallax measurements and their uncertainties. To a precision of 0.05 mas, we did not find a statistically significant offset in TGAS parallaxes (i.e., the global offset parameter $\varpi_{\rm 0} = 0.00\pm0.05$ mas) using our sample of distant RR Lyrae stars (median parallax of 0.8 mas and distance of 1.4 kpc). This result is consistent with the conclusion of \citet{cas16}, who use photometric parallaxes of distant Cepheids (median distance of $\approx2$ kpc) and find no offset to a precision of 0.02 mas.

The fact that we do not detect an offset in TGAS parallaxes may even be consistent with the findings of \citet{st16} and \citet{jao16}, who measure a global offset of $\approx-0.25$ mas in TGAS parallaxes, but suggest that it may become negligible for parallaxes smaller than a few mas (i.e., at large distances). However, given the manner in which the trigonometric parallax measurements are made (linear offsets on the detector), we do not understand what physical effect could cause a distance-dependent offset in TGAS parallaxes.

Regarding the uncertainty in TGAS parallaxes, we find no need to rescale formal parallax uncertainties {\em for RR Lyrae stars} (i.e., no need for $f_{\rm \varpi} > 1$), and recommend the following equation when calculating their uncertainty in parallax 
\begin{equation}
\sigma_{\rm \varpi, RRLyr} = \sqrt{(1.0\varsigma_{\rm \varpi})^2 + (0.18/1000)^2}\label{parallax_uncertainty},
\end{equation}
where the formal parallax uncertainty $\varsigma_{\rm \varpi}$ can be calculated using \autoref{reported_TGAS_par_unc}. The $f_{\rm \varpi} = 1.1$ and $\sigma_{\rm \varpi,add} = 0.12$ mas values obtained by \citet{gks16} are consistent at the $1\sigma$ level with our findings. 

We emphasize that our and \citet{gks16} results for $f_{\rm \varpi}$ and $\sigma_{\rm \varpi,add}$ were obtained using RR Lyrae stars. Due to as yet uncalibrated systematic effects in {\em Gaia} measurements, stars with different properties (e.g., color, brightness) may have different $f_{\rm \varpi}$ and $\sigma_{\rm \varpi,add}$ values. Various systematic effects may also explain the values of $f_{\rm \varpi} = 1.4$ and $\sigma_{\rm \varpi,add} = 0.20$ mas that \citet{gdr2} adopted for TGAS. Unlike us, \citet{gdr2} used a much more diverse sample of stars when constraining these two parameters (see their Appendix C.1).

By using individual measurements (e.g., TGAS parallaxes), PLZ relations, and parameters that model uncertainties in TGAS parallaxes, we have constrained distances to $\approx120$ RR Lyrae stars within 2.5 kpc of the Sun, to a 6\% precision. While this precision may seem quite low compared to precisions reported in some previous studies (e.g., a 0.8\% precision reported by \citealt{kle14}), we note that our estimate includes uncertainties due to correlations (e.g., between the period and metallicity slopes in the PLZ relation), and accounts for underestimated or unaccounted uncertainties in measurements and the model (via the $\sigma_{\rm M}$ parameter). Studies that did not take such uncertainties into account, and did not propagate them properly through the model, may have overestimated the precision of their distance measurements.

In conclusion, we are looking forward to applying the method and the experience developed in this work to the next Gaia data release, and doing some exciting Galactic science with Cepheid, Mira, and RR Lyrae stars in the near future.

\acknowledgments

B.S.~and H.-W.R.~acknowledge funding from the European Research Council under the European Union’s Seventh Framework Programme (FP 7) ERC Grant Agreement n.~${\rm [321035]}$. We thank Dr.~Adam G.~Riess for the thorough review, positive comments, and constructive remarks on this manuscript. This project was developed in part at the 2016 NYC Gaia Sprint, hosted by the Center for Computational Astrophysics at the Simons Foundation in New York City. This work has made use of data from the European Space Agency (ESA) mission Gaia (\url{http://www.cosmos.esa.int/gaia}), processed by the Gaia Data Processing and Analysis Consortium (DPAC, \url{http://www.cosmos.esa.int/web/gaia/dpac/consortium}). Funding for the DPAC has been provided by national institutions, in particular the institutions participating in the Gaia Multilateral Agreement. This publication makes use of data products from the Wide-field Infrared Survey Explorer, which is a joint project of the University of California, Los Angeles, and the Jet Propulsion Laboratory/California Institute of Technology, funded by the National Aeronautics and Space Administration.

\facilities{Gaia,WISE}




\bibliography{main}

\end{document}